\documentclass[a4paper,12pt]{article}

\usepackage{indentfirst}

\usepackage{amsfonts,amsmath,amssymb,bm,a4wide,graphicx,cite,makeidx,multicol}

\begin{document}

\title{Neutrino energy quantization in rotating medium}

\author{Alexander Grigoriev\footnote{ax.grigoriev@mail.ru},
\\
{\small {\it Skobeltsyn Institute of Nuclear Physics,}}
\\
   {\small {\it Moscow State University,}}
      {\small {\it 119992 Moscow,  Russia }}
      \\
Alexander Studenikin \footnote{studenik@srd.sinp.msu.ru },
   \\
   {\small {\it Department of Theoretical Physics,}}
   \\
   {\small {\it Moscow State University,}}
      {\small {\it 119992 Moscow,  Russia }}
}

\date{}
\maketitle

\sloppy

\begin{abstract}
Exact solution of the modified Dirac equation in rotating medium is found in polar coordinates in the limit of
vanishing neutrino mass. The solution for the active left-handed particle exhibit properties similar to those peculiar
for the charged particle moving in the presence of a constant magnetic field. Accordingly, the particle in the rotating
matter has circle orbits with energy levels, analogous to the Landau levels in magnetic field. The relevant physical
realization of the problem is motion of neutrino inside the rotating neutron star. The feature of such motion to form
binding states leads to the prediction of the new mechanism for neutrino trapping. The solution found can be used for
detailed description of relativistic and nearly massless neutrino dynamics in neutron stars. Results obtained further
develop the ``method of exact solutions" in application to particle interactions in presence of dense matter.
\end{abstract}

Recently a method for calculation of various processes of elementary particles interaction proceeding in matter has
been developed in a series of our papers (see \cite{StuJPA06_StuJPA08} and references therein). The framework of the
method is similar to the Furry representation in quantum electrodynamics and implies the use of exact solutions of the
wave equations for particles wave functions \cite{StuTerPLB05_GriStuTerPLB05,GriShiStuTerTroRPJ07_GriShiStuTerTroGC08}
to account for interaction with matter. The ``exact solutions method" \ has been already applied for description of
neutrino propagation in different media and electromagnetic fields and also for evaluation of the quantum theory of the
spin light of neutrino \cite{StuTerPLB05_GriStuTerPLB05} and spin light of electron
\cite{StuJPA06_StuJPA08,GriShiStuTerTroRPJ07_GriShiStuTerTroGC08} in matter, the two recently proposed new mechanisms
of electromagnetic radiation produced by a neutrino or an electron moving in matter.

The exact solutions of correspondent modified Dirac equations in
cases of a neutrino and an electron moving in matter at rest were
obtained in \cite{StuTerPLB05_GriStuTerPLB05} and
\cite{GriShiStuTerTroRPJ07_GriShiStuTerTroGC08} respectively. Then we
we have continued these studies and investigated neutrino behavior in
rotating medium with prospects for applications to neutron stars. We
have found \cite{GriShiStuTerTroRPJ07_GriShiStuTerTroGC08} (see also
the second paper of \cite{StuJPA06_StuJPA08}) the exact solution for
the neutrino wave function in the case of rotating medium using the
Cartesian coordinates. In this short note below we show how the same
problem can be solved using the polar coordinates. The obtained below
result contributes to the further development of the ``exact
solutions method" in application to studies of particle interactions
in presence of matter. Note that the employment of polar coordinates
in consideration of a neutrino motion in the rotating media fits the
symmetry of the problem, therefore the correspondent quantum numbers
that determines the neutrino quantum state have quite a natural
sense.

Let us consider unpolarized matter consisting of neutrons and
rotating around the third axes with the angular velocity $\omega$.
For definiteness, we consider below the case of an electron
antineutrino motion in such a medium. The problem for different other
neutrino species can be solved in a similar way.

Our starting point is the modified Dirac equation
\cite{StuTerPLB05_GriStuTerPLB05} accounting for the neutrino
interaction with matter,
\begin{equation}\label{Dirac} \Big\{
\gamma_{\mu}p^\mu-\frac{1}{2} \gamma_{\mu}(1+\gamma_{5})f^{\mu}-m \Big\}\Psi(x)=0,
\end{equation}
where
\begin{equation}\label{f}
  f^\mu={G_F \over \sqrt2}(1+4\sin^2 \theta_W) j^\mu=\widetilde{G}_F
  j^\mu.
\end{equation}
The matter current is given by
\begin{equation}
j^\mu=(n,n{\bf v}) \label{j}
\end{equation}
where $n$ is the invariant matter density and ${\bf v}$ being the
macroscopic matter speed. In our particular case the matter current
can be written as $j^\mu=n(1,-\omega y,\omega x,0)$, where $y$ and
$x$ are the spatial coordinates in the plane orthogonal to the
rotation axis.

Introducing the notation $\gamma =\widetilde{G}_Fn\omega$ and using
polar coordinates we rewrite the equation (\ref{Dirac}) in
components:
\begin{equation}
\label{DiracComponents}
\left\{
\begin{array}{rcl}
  m\Psi_{1,2} +ie^{\mp i\varphi}[\frac{\partial}{\partial r}
  \mp \frac{i}{r}\frac{\partial}{\partial \varphi} \pm \gamma r]\Psi_{4,3} -(E+\widetilde{G}_F \pm p_3)\Psi_{3,4}=0 \\
  \vspace{-0.2cm} \\
  (E \mp p_3)\Psi_{1,2} +ie^{\mp i\varphi}[\frac{\partial}{\partial r}
  \mp \frac{i}{r}\frac{\partial}{\partial \varphi}]\Psi_{2,1} -m\Psi_{3,4}=0 \\
\end{array}
\right.
\end{equation}
where $p_3$ is the neutrino momentum operator. The chiral
representation of Dirac matrices is used. The form of the equations
implies the following operators to be integrals of motion: the energy
$E=i\partial_t$, the third momentum component $p_3=-i\partial_3$, and
the third component of the total angular momentum (which is the sum
of the orbital and spin momenta, $J_3=L_3+S_3=
-i\partial_{\varphi}+\Sigma_3$). Therefore we write the solution
\begin{equation}\label{ansatz1}
        \Psi ({\mathbf r}, t)
        =\displaystyle{e^{-iEt+ip_3z+i(l-1/2)\varphi}}{\psi},
\end{equation}
where the function $\psi$ can be written in the following form
\begin{equation}\label{ansatz2}
    {\psi}=
        \left(%
\begin{array}{lc}
  \psi_1 e^{-i\varphi} \\
  \psi_2 e^{i\varphi} \\
  \psi_3 e^{-i\varphi} \\
  \psi_4 e^{i\varphi} \\
\end{array}
\right).
\end{equation}
Substituting relations (\ref{ansatz1}) and (\ref{ansatz2}) into the
system (\ref{DiracComponents}) one finds the following set of
equations for the functions ${\psi}_i$:
\begin{equation}\label{DiracComponents_reduced}
\left\{
    \begin{array}{rcl}
  \displaystyle m{\psi}_1 -(E+p_3+\widetilde{G}_Fn){\psi}_3
  +i\sqrt{\gamma}\left[\frac{\partial}{\partial r} +\frac{l}{r} +r\right]{\psi}_4 =0 \hphantom{.}
  \\
  \displaystyle m{\psi}_2
  +i\sqrt{\gamma}\left[\frac{\partial}{\partial r} -\frac{l-1}{r} -r\right]{\psi}_3
  -(E-p_3+\widetilde{G}_Fn){\psi}_4 =0 \hphantom{.}
  \\
  \displaystyle (E-p_3){\psi}_1
  +i\sqrt{\gamma}\left[\frac{\partial}{\partial r} +\frac{l}{r} \right]{\psi}_2 -m{\psi}_3 =0 \hphantom{.}
  \\
  \displaystyle i\sqrt{\gamma}\left[\frac{\partial}{\partial r} -\frac{l-1}{r} \right]{\psi}_1 +(E+p_3){\psi}_2 -m{\psi}_4 =0 .
  \end{array}
\right.
\end{equation}
In the case of relativistic neutrinos (i.e., in the case of vanishing
neutrino mass) the fist and the second pairs of the equations
decouple from each other and describe, respectively, active
left-handed ($L$) and sterile right-handed ($R$) neutrino states.

Now let us turn to the first pair of equations in
(\ref{DiracComponents_reduced}). Expressing one wave function
component (${\psi}_3$ or ${\psi}_4$) through another and introducing
the substitution $\rho =\gamma r^2$ we obtain separate equations for
the each of the components,
\begin{equation}\label{Equations f_3,f_4}
\begin{array}{c}
  \displaystyle \left( \rho\frac{\partial^2}{\partial \rho^2} +\frac{\partial}{\partial \rho} -\frac{(l-1)^2}{4\rho} -\frac{l}{2}
    -\frac{\rho}{4} +\frac{(E+\widetilde{G}_Fn)^2-p_3^2}{4\gamma} \right){\psi}_3 =0, \\
  \displaystyle \left( \rho\frac{\partial^2}{\partial \rho^2} +\frac{\partial}{\partial \rho} -\frac{l^2}{4\rho} -\frac{l-1}{2}
    -\frac{\rho}{4} +\frac{(E+\widetilde{G}_Fn)^2-p_3^2}{4\gamma} \right){\psi}_4 =0.\\
\end{array}
\end{equation}
These  equations are almost identical to those corresponding to the
case of charged particle motion in the constant homogeneous magnetic
field when the problem is considered in the frame of polar
coordinates \cite{SokTerSynRad68}. Their solutions, that posses
physical meaning, are expressed via the Laguerre functions $I_{N, \
s}$ with $N$, $s=N-l$ and $l$ being the principal, radial and total
orbital momentum quantum numbers. The energy-momentum relation
therewith is expressed by the formula
\begin{equation}
\label{Spectr} E_N = \sqrt{p_3^2 + 2 N \gamma} + \widetilde{G}_Fn.
\end{equation}

Let us consider the second equation of the system
(\ref{DiracComponents_reduced}). Presenting the functions
${\psi}_{3,4}$ in the form
\begin{equation}\label{f_3,4}
    {\psi}_3=C_3 I_{N-1,\ s}(\rho), \ \ \ {\psi}_4=C_4 I_{N, \ s}(\rho).
\end{equation}
and taking into account the relation
\begin{align}\label{relation f3_f4}
    i\sqrt{\gamma}\left[\frac{\partial}{\partial r} +\frac{l}{r} +r\right]I_{N-1, \ s}(\rho)= & \notag \\
    i\sqrt{\gamma \rho}\left[ 2\frac{\partial}{\partial \rho}+1+ \frac{l}{\rho}\right]&
    I_{N, \ s}(\rho)=i\sqrt{2N\gamma}I_{N-1, \ s}(\rho)
\end{align}
we obtain relation between coefficients $C_3$ and $C_4$, that leads
to
\begin{equation}\label{ratio}
    \frac{C_3}{C_4}=-i\frac{E-p_3+\widetilde{G}_F}{\sqrt{2n\gamma}}.
\end{equation}
The remaining non-defined coefficient is determined by the
normalization condition for the wave function. Finally for the
relativistic active left-handed antineutrino wave function we get
\begin{align}\label{solution}
     & \Psi_L= \\
     & \frac{e^{-iEt + ip_3z} \ e^{i(l-1/2)\varphi}}
       {\sqrt{2\pi L}\sqrt{(E+\widetilde{G}_Fn-p_3)^2+2N\gamma}}\left(
\begin{array}{c}
0 \\
0 \\
(E+\widetilde{G}_Fn-p_3) \ I_{N-1, \ s}(\rho)\ e^{-i\varphi} \\
i\sqrt{2N\gamma} \ I_{N, \ s}(\rho) \ e^{i\varphi}
\end{array}
\right), \notag
\end{align}
where $L$ in the left-hand side is the normalization length.

The solution for the relativistic sterile right-handed  neutrino
state can be presented in the plane-wave form
\begin{equation}
\label{PsiR} \Psi_R = \frac{e^{-i p
x}}{L^{3/2}\sqrt{(p_3-E)^2+p_1^2+p_2^2}}\left(
\begin{array}{c}
0 \\
0 \\
p_1 - i p_2 \\
p_3 - E
\end{array}
\right)
\end{equation}
with the vacuum energy-momentum relation $E=p$.

As it follows from (\ref{Spectr}), we conclude that the transversal
motion of an active neutrinos and antineutrinos is quantized in
moving matter \cite{GriSavStuRPJ07} very much like an electron energy
is quantized in a constant magnetic field that corresponds to the
relativistic form of the Landau energy levels (see, for instance,
\cite{SokTerSynRad68}). Consider again antineutrino. The transversal
motion momentum is given by
\begin{equation}
\tilde p_{\bot}= \sqrt{2\rho N}.
\end{equation}
The quantum number $N$ determines also the radius of the antineutrino
quasi-classical orbit in matter (it is supposed that $N\gg 1$ and
$p_3=0$),
\begin{equation}\label{R}
    R=\sqrt{\frac{2N}{\widetilde{G}_Fn\omega}} \ .
\end{equation}
From this it follows, for instance, that low energy (but still
relativistic) antineutrinos can have bound orbits inside a rotating
star. This can lead to the mechanism of low-energy neutrinos trapping
inside rotating neutron stars.

\end{document}